\documentclass[aps,prl,preprint,groupedaddress]{revtex4-2}
\usepackage{graphicx}
\usepackage{amsmath}
\begin{document}


\title{Wall accumulation of confined active Janus colloids due to effective active diffusivity}


\author{Sandeep Ramteke}
\affiliation{Dept. of Mechanical Engineering, Florida International University}
\author{Alicia Boymelgreen}
\affiliation{Dept. of Mechanical Engineering, Florida International University}
\author{Jarrod Schiffbauer}
\affiliation{Dept. of Physical and Environmental Sciences, Colorado Mesa University}
\affiliation{Dept. of Mechanical Engineering, University of Colorado at Boulder}

\date{\today}

\begin{abstract}
Electrokinetically-driven Janus colloids, e.g., with one metallic and one dielectric hemisphere, confined between parallel walls exhibit a boundary-accumulation mechanism enabled by an effective cross-channel diffusivity which is distinct from wall accumulation of active Brownian or run-and-tumble particles. Using density-matched suspensions and three-dimensional confocal imaging, we directly measure the full time-dependent redistribution of particles across the channel under an applied AC electric field. The wall population grows exponentially while the bulk depletes, and data obtained over multiple field strengths collapse onto a single curve when rescaled by the measured relaxation rate, revealing one dominant, confinement-controlled timescale. Propulsion follows the expected induced-charge electrophoretic scaling, with a mean orientation angle lying between $2^{\circ}$ and $10^{\circ}$ above horizontal, leading to a top-biased accumulation. Comparison with an overdamped Ornstein-Uhlenbeck turning model suggests that persistent stochastic turning about a small out-of-plane angle results in a cross-channel effective drift and diffusion. The drift governs the dominant timescale and the diffusion is strong enough to provide significant accumulation on the bottom wall despite a mean upward orientational bias.
\end{abstract}


\maketitle

\begin{figure}
\centering
 \includegraphics[width=0.8\columnwidth]{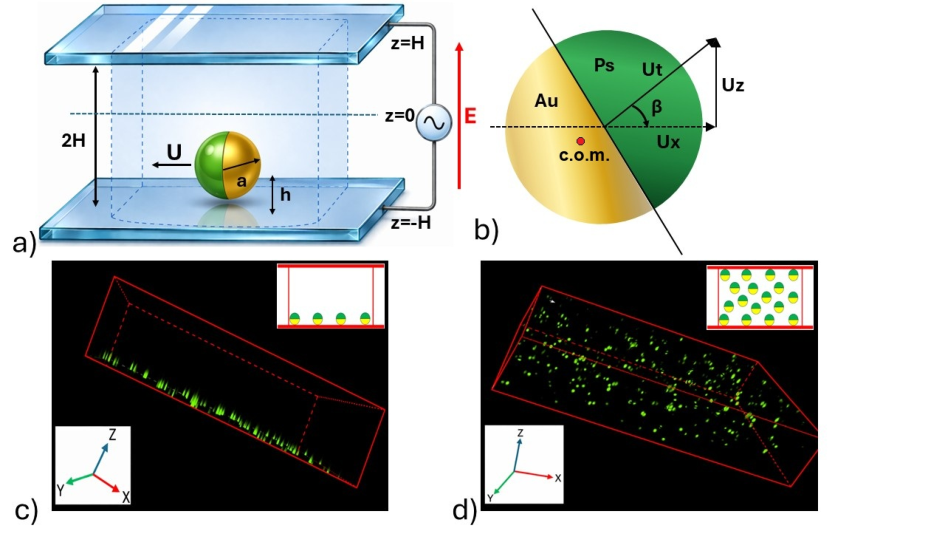}
 \caption{\label{fig1}a) Schematic of experimental set up electric field driven Janus particle in microfluidic chip. b) Schematic of Janus particle with center of mass indicated by a red dot at orientation angle $\beta$ with velocity components in x and z directions represented respectively by $U_x$ and $U_z$. c-d) Z- scan of JP distribution in c) pure DI water (sedimentation dominated) and d) 0.70 w/w glycerol /DI (density matched)}
 \end{figure}

\begin{center}
\begin{figure}[h]
\includegraphics[scale=1.0]{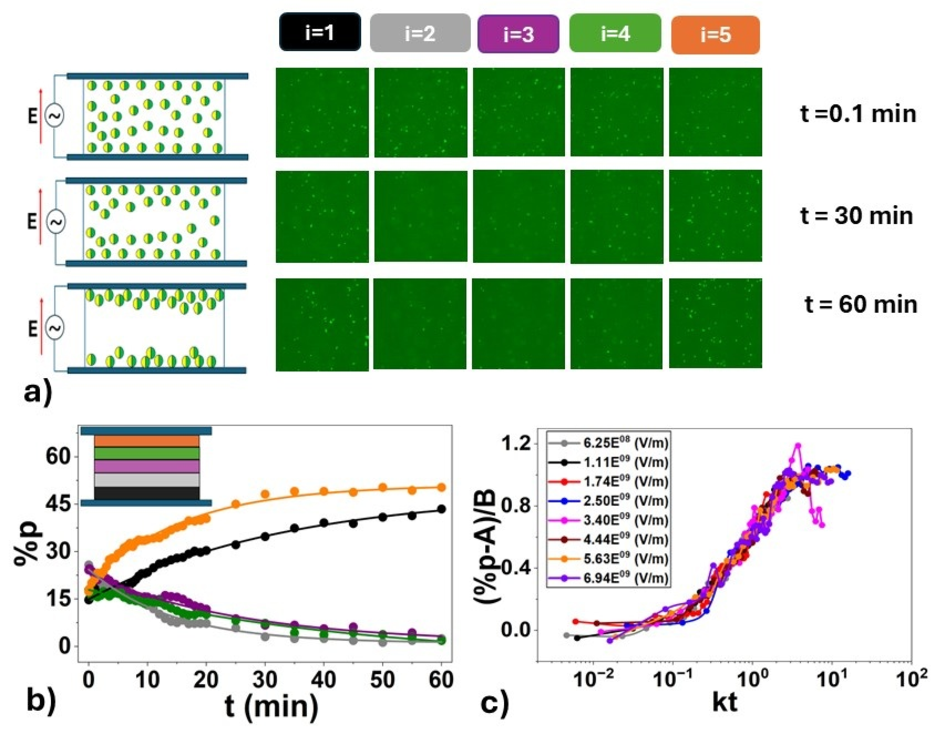}
\caption{\label{fig2}a) Schematic illustration and corresponding fluorescence images showing progressive accumulation time scale for particles in 5 layers of thickness $2H/5$, (black ($i=1$), gray ($i = 2$), magenta ($i = 3$), green ($i = 4$), orange ($i = 5$). b) Percentage of Janus particles located in each layer as a function of time in each layer at $8 \textrm{V}$, corresponding to an RMS field of $4.72\times 10^4 \enspace \textrm{V}/\textrm{m}$  c) Accumulation profile evaluated in the bottom layer $i = 1$ for varying voltages collapse when scaled with non-dimensional time, $kt$.}
\end{figure}
\end{center}

\begin{center}
\begin{figure}
 \includegraphics[width=0.8\columnwidth]{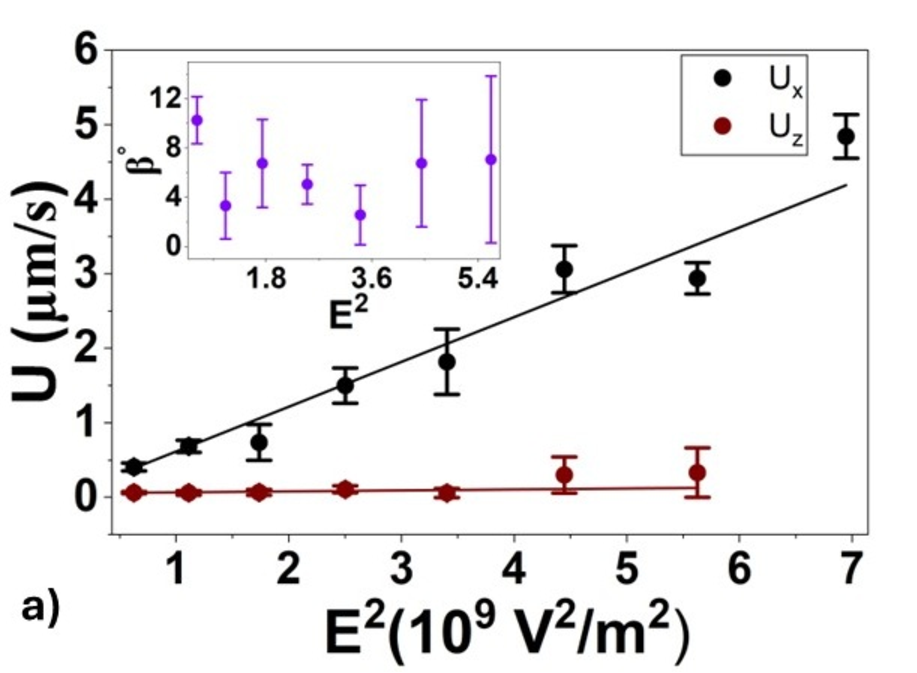}
 \includegraphics[width=0.8\columnwidth]{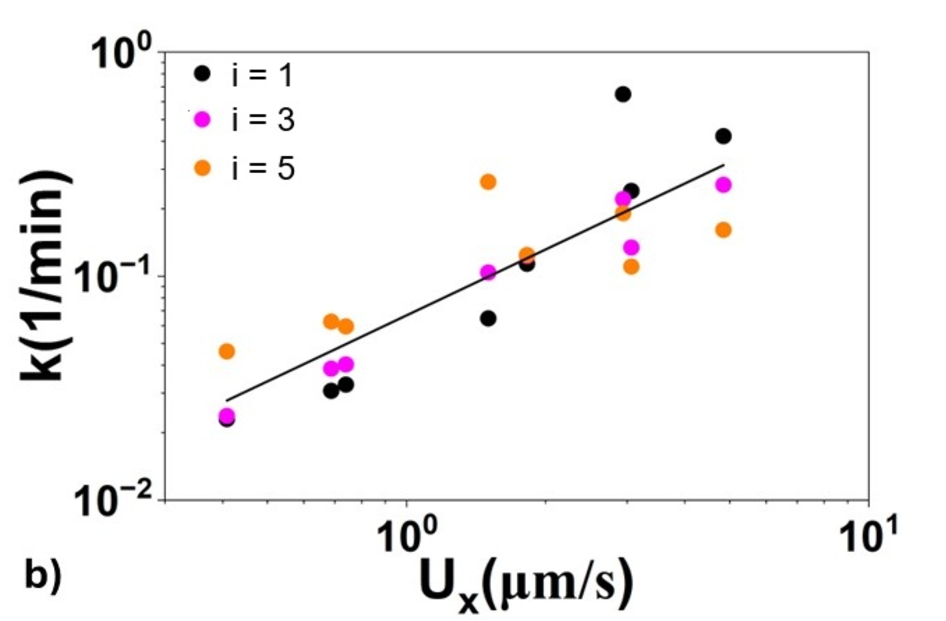}
 \caption{\label{fig3}a) Velocity scaling of Janus particles with electric field strength. The black dots represent the Velocity $U_x$ in x direction, the black line validates the quadratic scaling expected in ICEP. The brown dots represent the measured velocity $U_z$ in z direction. b) Accumulation rate constant $k$ vs particle velocity $U_x$. Colors correspond to layers; red ($i=1$), Blue ($i=3$), pink ($i=5$) }
 \end{figure}
 \end{center}

\begin{figure}
 \centering
 \includegraphics[width=0.8\columnwidth]{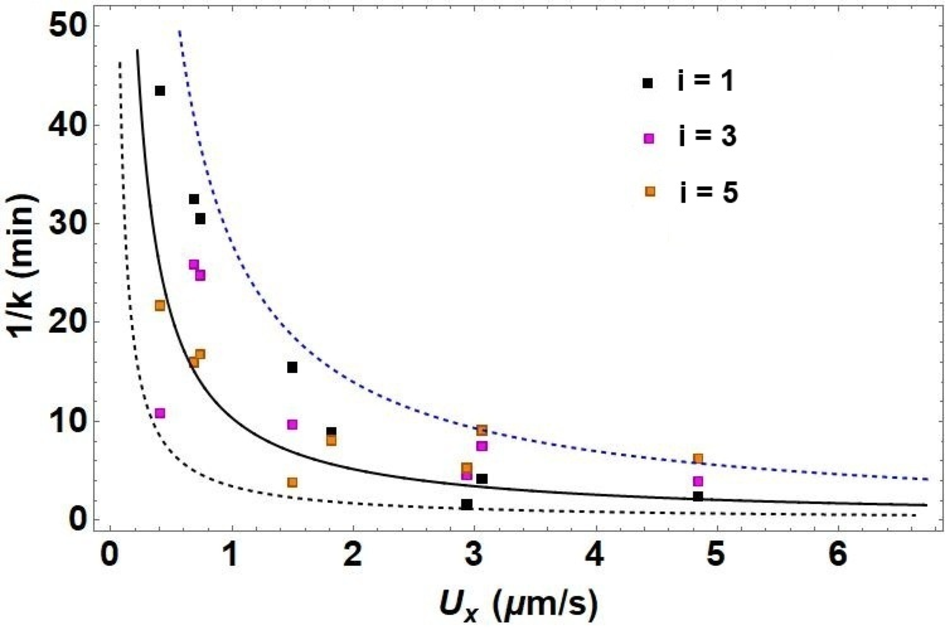}
 \caption{\label{fig4} Experimental $1/k$ (blocks) and theoretical $1/k_{1,min}$ values (lines) vs. $U_x$ are shown. The experimental legend is the same as Fig. 3b and the solid black line is calculated using $\beta = 2.7^{\circ}$, while the minimum observed experimental value, $\beta = 2.^{\circ}$ is dashed blue and maximum value, $\beta = 10^{\circ}$ is dashed black.}
 \end{figure}

The orientational dynamics of active, low Reynolds micro swimmers\cite{purcell1977} are often framed in terms of one of two canonical models: active Brownian particles (ABPs), whose propulsion direction evolves through continuous rotational diffusion\cite{ABP1,ABP2}, and run-and-tumble (RT) swimmers, whose persistent runs are interrupted by discrete reorientation events\cite{RT1,RT2}. When confined, both are known to be attracted to and accumulate at boundaries\cite{elgeti2013,spagnolie2012,sindelka2025,baouche2026,shaik2023}. Realistic active colloids typically have an asymmetric density distribution leading to an off-center center of mass (see Fig.\ref{fig1}b). Hence, they fall outside both standard pictures due to gravity-induced orientational dynamics\cite{Zeng2022,Das2020} which manifest in a biased wall accumulation under density matched conditions\cite{Carrasco2023}.\\
\indent The translational diffusion coefficient, $D_T$, for micron-sized spherical particles in water or water-glycerol mixtures is typically in the range of $10^{-15}$ to $10^{-14} \quad m^2/s $. Consequently, the trajectories of individual JPs appear essentially ballistic over typical experimental time and length scales. However, consider spherical particles of radius $a$ confined to a cell with a total height $2H$, as in Fig.\ref{fig1}. The translational diffusion time corresponding to diffusion across the system in the z-direction is $\tau_{T}= 4H^2/D_T$, while the rotational diffusion time is $\tau_{rot} = 1/6D_{R}$ for rotational diffusivity $D_R$. Assuming the usual Stokes-Einstein relations for a spherical particle, if $a^2/18 H^2 \ll 1$, rotational diffusion can affect the overall particle flux and long-time distribution even when the observable effect on a single particle trajectory is very subtle.\\
\indent In the present work we employ density-matched suspensions of $5 \enspace \mu \textrm{m}$ diameter gold-polystyrene JPs using glycerol mixtures (DI-glycerol)\cite{supmat} to suppress gravitational sedimentation and enable the study of particle redistribution in the full three-dimensional volume of the chamber. Critically, however, this does not eliminate the gravitational torque on the off-center JP center of mass. Hower, the application of a uniform AC electric field polarizes the metal/dielectric hemispheres asymmetrically and drives induced charge electrophoretic (ICEP) propulsion approximately transverse to the field\cite{squires2006} with a speed $\mu_{ICEP}E^2$ where $\mu_{ICEP}$ is the ICEP mobility and $E$ the magnitude of the field. The same electrohydrodynamic response plus residual, competing gravitational and electric torques selects an off-horizontal preferred equilibrium angle\cite{boymelgreen2024,sofer2023}. Thus, in a viscous fluid, any perturbations away from this orientational equilibrium relax continuously rather than through isotropic free rotational diffusion or discrete tumbles\cite{ABP2,RT2}.\\
\indent This orientational physics has direct consequences for transport in confinement\cite{khatri2026}. EK-JPs do not execute a standard persistent random walk. They instead exhibit persistent stochastic turning about a preferred pitch while undergoing more or less ballistic translation along the $x-y$ plane with a constant propulsion speed. This results in out-of-plane, quasi-helical trajectories with a distribution of net $z$ displacements. As a result, particles accumulate preferentially at the upper wall while still exhibiting substantial accumulation at the lower wall.\\
\indent Experimental visualization is conducted using a spinning disc confocal microscope with a motorized stage and a 20x objective. The experimental chamber consists of a silicon reservoir ($2H=120 \enspace \mu \textrm{m}$) sandwiched between a single Indium tin oxide (ITO) coated glass slide and an ITO coated cover slip which act as conducting electrodes.  The ITO electrodes are connected to an AC function generator supplying voltages between 5 VPP to 20 VPP at a frequency of 1 kHz (see Fig.\ref{fig1}a and \cite{supmat}).  A point-by-point z-scan is used to automatically image the distribution of particles in 5 equally-spaced planes across the reservoir.\\
\indent Figure \ref{fig1}a illustrates the experimental platform used to study wall accumulation of EK-JPs along with the relevant geometry (Fig.\ref{fig1}b). In pure DI water, JPs sediment rapidly and collect near the bottom surface due to gravity(Fig.\ref{fig1}c). In contrast, in the density-matched solution, particles remain distributed throughout the chamber, suppressing sedimentation and enabling three-dimensional transport measurements(Fig.\ref{fig1}d).\\
\indent The stability of the density matched solution is verified by observing the suspension for 1 hour (see\cite{supmat}). After applying the field, the resulting distribution is observed at discrete time intervals across the five z layers over an hour (Fig. \ref{fig2}a). Figure \ref{fig2}b quantifies the field-driven redistribution of JPs, with accumulation at the confining walls and depletion of the bulk (near center) region. Measurements for applied voltages in the range $ V= 6-20\textrm{V}$ are provided in \cite{supmat}. Recalling $U\propto E^2$, we verify the growth rate is exponential and scales with velocity by measuring the accumulation at different voltages as shown in Fig. \ref{fig2}c. Rescaling time by the fitted rate collapses the wall-accumulation curves across field strengths, indicating one dominant confinement-controlled redistribution timescale, $1/k$. \\
\indent The temporal evolution of the particle fraction observed in the $i^{th}$ focal layer, $p_i(t)$, was fit to a first order exponential growth model given by
\begin{equation}p_i(t)=A+B(1-\exp{-kt})\end{equation}
where $A$ is the initial particle population at time $t=0$, $B$ is the accumulation of particles in steady state, and $k$ is an accumulation rate constant. Thus, the saturation value is given by $A+B$ at $t\to\infty$. For each applied voltage, the time dependent population was extracted for all five layers ($i=1$, $2$, $3$ , $4$, $5$). Accumulation was evident in the top ($i=5$) and bottom ($i=1$) layers, had $B>0$ values while the bulk-like middle layers ($B<0$) underwent depletion.\\
\indent Figure \ref{fig3} a shows that the propulsion of the EK-JP particle given by $U_x$ increases approximately linearly with $E^2$, confirming the expected induced-charge electrophoresis (ICEP) scaling where particle speed is proportional to square of applied electric field. The out of plane velocity $U_z$ ranges between $1.73\times 10^{-3} \enspace \mu \textrm{m}/\textrm{s}$ and $6.5\times 10^{-2} \enspace \mu\textrm{m}/\textrm{s}$. The ratio of velocity components is found to be in the range $0.088 < U_z/U_x  < 0.18$, giving corresponding (time-averaged) orientation in the range $ 2^{\circ} < \beta <10^{\circ}$ where $\beta =  \arctan{U_z⁄U_x}$ as shown in the inset of Figure \ref{fig3}a. Since $U_z=U(E) \sin{\beta(E,t)}$ and fluctuations and variations in $\beta(E,t)$ have an unclear dependence on $E$, $U_z$ is observed to deviate from the $E^2$ scaling\cite{supmat}. Note that particle distribution rate (Fig. \ref{fig3}b) increases with $U_x$, indicating the faster swimmers accumulated and depleted more rapidly. However, $k$ values extracted in the regions showed no distinct trend in $z$ (Fig. \ref{fig3}b), suggesting a single dominant mode is observed across the channel. The linear trend suggests that the accumulation rate of the particles is governed by the propulsion driven transport.\\
\indent The theoretical model assumes that the system has translational invariance in the $x-y$ plane and that the low-Reynolds motion is overdamped. For simplicity, we take the system to be essentially two dimensional, confined by no-flux boundaries at $z=\pm d$ (where $d=H-a$ is the accessible half-channel height), and subject to an alternating electric field of magnitude $E$ in the z-direction so that the preferred propulsion direction of particles is more or less horizontally-directed at some small angle $\beta_*$ from the x-axis with speed $U(E)=\mu_{ICEP}E^2$. At this point, we allow for the possibility that $\beta_*(E)$ \cite{boymelgreen2024}, but do not assume any specific dependence. To capture the experimentally observed late-time exponential bulk redistribution, $\beta$ is modeled as a linear mean-reverting stochastic process and coarse-grained over in the fast-relaxation limit. Then, the lowest eigenmode is obtained for the average density distribution in the plane, $n(z,t)=\int_{-\pi}^{\pi} P(z,\beta,t)d\beta$ and compared to the observed depletion rates. Here $P(z,\beta,t)$ is the solution to the Fokker-Planck equation associated with the equations of motion,
\begin{equation}\label{lan1}\dot{\beta}=-\lambda(E)(\beta-\beta_*(E))+\sqrt{2D_R}\xi_{\beta}(t)\end{equation}
and
\begin{equation}\label{lan2}\dot{z}=U(E)\sin{\beta(E)}+\sqrt{2D_T}\xi_z (t) \end{equation}
where $\xi_{z,\beta} (t)$ are independent white Gaussian noises and $\beta(t)=\beta_*+\delta\beta(t)$ assuming both $\beta$ and $\delta \beta \ll 1$. The (possibly field-dependent) angular relaxation, $\lambda(E)$ corresponds to an orientational relaxation time, $\tau_{or}=1/\lambda$, distinct from $\tau_{rot}$. Here the linear dependence in angular displacement is generally applicable provided the assumption that both the angle and angular fluctuations are small is valid. The fact that the system is driven back continuously towards a preferred orientation while subject to thermal fluctuations in the velocities, $\dot{\beta}$ and $\dot{z}$, results in colored noise fluctuations $\delta\beta(t)$ with variance $D_R/\lambda(E)$, which augments the drift in the z-direction. Provided we restrict our comparison to cases where the relevant time $t\gg\tau_{or}$, the angular coarse-graining is justified. Accordingly, effectively averaging over the $x-y$ plane and the full distribution of angles about $\beta_*$, the wall accumulation problem can be characterized by an effective drift $V_{eff}(E)=U(E)\beta_*(E)$ and an effective diffusivity, $D_{eff}=D_T + U(E)^2D_R/\lambda(E)^2$, where the second term represents the active contribution to the diffusion. The lowest eigenmode for the solution $n(z,t)$ is given as $k_1=\pi^2 D_{eff}/4d^2 + V_{eff}^2/4 D_{eff}$. Additional details are provided in the Appendix\ref{app}. Since the viscosity of the density-matched suspension is quite high and corresponding $D_T$ quite low, the system dynamics occupy a regime where $D_{eff}\sim U(E)^2D_R/\lambda(E)^2$.\\
\indent In the absence of a specific model $\lambda(E)$, we can minimizes the growth rate with respect to $\lambda$ as an estimate. This gives $k_{1,min}=5\pi U(E)\beta_*(E)/8d$, the inverse of which is plotted in Fig.\ref{fig4} along with experimentally obtained $1/k$ depletion times taken from bulk (middle 3) layers, top, or bottom as indicated\cite{supmat}. Here, $U(E)$ is determined using the mobility, $\mu_{ICEP}= 1.55\times10^{-16} \textrm{m}^3/\textrm{V}^2 \textrm{s}$ \cite{supmat}. Since there isn't a clear trend with $\beta(E)$ in the present data (see inset of Fig.\ref{fig3}a,) $\beta_*$ is taken as the average of all experimental values and minimum and maximum bounds are indicated. Despite the agnosticism regarding $\beta_*(E)$ and $\lambda(E)$, the $k_{1,min}$ model is in reasonably good agreement with the experimentally observed timescale.\\
\indent In summary, the results here have demonstrated that density-matched, electrokinetically driven Janus swimmers confined in a microfluidic chamber undergo robust time-dependent wall accumulation governed by persistent stochastic turning rather than conventional ABP or RT dynamics. The measured propulsion follows the expected ICEP scaling, $U_x\propto E^2$, while perturbations to a small but finite out-of-plane component generate effective cross-channel drift and diffusion. This effective diffusion accounts for the net downward transport of a significant fraction of particles even when the mean bias is upward. Thus, density-matched EK-JPs stand in contrast to strongly gravitactic bottom-heavy Janus swimmers that are dominated by pronounced top-wall trapping\cite{campbell2013,Das2020,Carrasco2023}. The resulting temporal evolution of the layer populations is well described by a single exponential relaxation with rate constant $k\propto U_z$. The reduced model captures this late-time relaxation scale but does not account for the near-complete depletion observed experimentally. That stronger depletion likely reflects additional boundary physics, such as wall residence or trapping effects\cite{spagnolie2012,vanBaalen2025}, which are not included in the present minimal description. Nonetheless, the resulting top-biased yet distinctly two-wall accumulation demonstrates that, even in a density-matched Newtonian system, weakly biased reorientation in confinement can govern macroscopic active transport beyond standard isotropic swimmer models.\\

\appendix*\label{app}
\section{Appendix: Details on theoretical model}
Key steps in the derivation of the growth rate are given below. Starting with the Langevin equations for $\beta$ and $z$ defined in the main text and the autocorrelation for $\delta \beta$,
\begin{equation} \langle \delta \beta(t) \delta \beta(0) \rangle = \frac{D_R}{\lambda(E)}\exp{-\lambda(E)t}, \end{equation}
the velocity in the z-direction for small $\beta$ is given by $U_z(E) = U(E)\beta(E)$, which is essentially a constant drift, $V_{eff} = U(E)\beta_*(E)$, plus a fluctuation, $\delta V(t) = U(E)\delta \beta$, which can be associated with the effective active diffusion in the z-direction,
\begin{equation}D_{act}=\int_{0}^{\infty} dt \langle \delta V(t) \delta V(0) \rangle =\frac{ U(E)^2 D_R}{\lambda(E)^2}. \end{equation}
Hence, with the corresponding Fokker-Planck equation,
\begin{equation}
\begin{split}
\partial_t P  &= -\partial_z (U(E)\beta(E)P) + D_T\partial_z^2 P \\
  &+\partial_{\beta}[\lambda(E)(\beta-\beta_*)P + D_R\partial_{\beta}P]
\end{split}
\end{equation}
and no-flux boundary conditions at $z=\pm d$, we can define a coarse-grained $n(z,t)=\int P(z,\beta,t) d\beta$, which obeys a 1D effective drift-diffusion equation,
\begin{equation}\partial_t n=-V_{eff}\partial_z n + D_{eff}\partial_z^2 n \end{equation}
with no flux conditions, $V_{eff}n-D_{eff}\partial_z n = 0$ at $z=\pm d$ and a total effective diffusivity, $D_{eff}= D_T + D_{act}$. This has solution,
\begin{equation}n(z,t) = n_{ss}(z) + \sum_{m=1}^{\infty}A_m \phi_m(z) e^{-k_m t} \end{equation} where the steady-state solution is $n_{ss}(z) = C\exp{(V_{eff} z/D_{eff})}$ and the growth rate for the $m$th eigenfunction $\phi_m (z)$ is
\begin{equation}k_m = D_{eff}(\frac{m \pi}{2d})^2 +\frac{V_{eff}^2}{4 D_{eff}} \end{equation}

\begin{acknowledgments}
The authors would like to acknowledge financial support for this research from award numbers 2126451, 2430509 from the National Science Foundation and award number 80NSSC24K0415 from NASA Science Mission Directorate.
\end{acknowledgments}

\bibliography{turning}

\end{document}